\documentclass[aps,prd,nofootinbib,superscriptaddress,floatfix,nofootinbib,amsmath,amssymb]{revtex4}

\usepackage[dvips]{graphicx}

\usepackage[USenglish]{babel}
\usepackage{amsmath,amssymb,amsxtra,amsfonts}
\usepackage{graphicx}
\usepackage{dcolumn}
\usepackage{bm}
\usepackage{amssymb}
\usepackage{latexsym}
\usepackage[colorlinks, linkcolor=blue, citecolor=blue, urlcolor=blue]{hyperref}

\newcommand{\be}{\begin{equation}}
\newcommand{\ee}{\end{equation}}
\newcommand{\bq}{\begin{eqnarray}}
\newcommand{\eq}{\end{eqnarray}}

\begin{document}
\bibliographystyle{plain}

\begin{flushright}
BNU
\end{flushright}

\title{Distinguishing between the inhomogeneous model and $\Lambda$CDM model with cosmic age method}

\author{Siqi Liu}
\email{tjzhang@bnu.edu.cn} \affiliation{Department of Astronomy, Beijing Normal University， Beijing 100875, China}

\author{Tong-Jie Zhang}
\email{tjzhang@bnu.edu.cn} \affiliation{Department of Astronomy, Beijing Normal University， Beijing 100875, China} \affiliation{Center for High Energy Physics, Peking University, Beijing 100871, China}

\begin{abstract}
Cosmological observables could be used to construct cosmological models; however, a fixed number of observables limited to the light cone are not enough to uniquely determine a certain model. In this paper, we employ a reconstructed spherically symmetric, inhomogeneous model that shares the same angular-diameter-distance-redshift relationship $d_A(z)$ and Hubble parameter $H(z)$ besides $\Lambda$CDM model (which we call LTB-$\Lambda$CDM model in this paper), that may provide another solution. Cosmic age, which is off the light cone, could be used to distinguish between these two models. We derive the formulae for age calculation with origin conditions. From the data given by 9-year WMAP  measurement, we compute the likelihood of the parameters in these two models respectively by using the Distance Prior method and perform likelihood analysis by generating Monte Carlo Markov Chain for the purpose of bringing tighter constraints on the parameters $\Omega_m$ and $H_0$ (the parameters that we use for calculation). The results yield: $t_{\Lambda\textrm{CDM}} =13.76 \pm 0.09 ~\rm Gyr$, $t_{\rm {LTB}-\Lambda\textrm{CDM}} =11.38 \pm 0.15 ~\rm Gyr$, both in $1\sigma$ agreement with the constraint of cosmic age given by metal-deficient stars.
\end{abstract}

\maketitle

\section{Introduction}\label{sec:intro}
In the past decades, remarkable progress has been made in measuring cosmological parameters to unprecedented accuracy. Data of the Cosmic Microwave Background Radiation(CMBR)\cite{WMAP9} indicates a flat Universe, and the observations of type \uppercase \expandafter {\romannumeral 1}a supernovae (SNe \uppercase \expandafter {\romannumeral 1}a)\cite{SNR}\cite{SNP} show that the universe is undergoing an accelerated expansion. The matter content of the universe falls well short of the energy density necessary to provide a flat curvature in the standard cosmological model which is homogeneous and isotropic, the Friedmann-Lema\^{\i}tre-Robertson-Walker (FLRW) metric. The most popular interpretation of this mismatch is that the ‘missing’ density is assumed to be present in the form of dark energy which provides a pressure leading to the acceleration of the universe.

The observations that lead to the assumptions of dark energy are true, but do not necessarily imply that dark energy as usually envisioned indeed exists. Another theoretical approach was raised to explain the current cosmic observations, in which there is no need for dark energy, no need for new long-range forces or modifications of general relativity, new ultra-light particles or anthropic reasoning. We follow the prescription of \cite{MHE}, applied to $\Lambda$CDM model by \cite{CBKH} and explore the implications of the fact that one can construct a spherically symmetric inhomogeneous model that exactly reproduces the angular-diameter-distance-redshift relationship $d_A(z)$ and Hubble parameter $H(z)$ of any given observations based on $\Lambda$CDM model, which we call \emph{LTB-$\Lambda$CDM model} in this paper.

As $\Lambda$CDM and LTB-$\Lambda$CDM models share the same observables limited on the light cone, we can only distinguish these two models by other physical quantities off the light cone. The age of the Universe, which satisfies this requirement, can be a good choice to break the degeneracy. The method to calculate the age of the universe at $r =0$ (corresponding to our position in the universe) in LTB-$\Lambda$CDM model is based on the original conditions of spherical coordinate.

To calculate the age of the universe, we use the parameters given by 9-year WMAP. However, directly using the WMAP results for $\Omega_m$ and $H_0$ overestimates the uncertainty, due to the fact that these two parameters are not orthogonal, which means their uncertainties are correlated. Instead, in \cite{WMAP5} there is a method called Distance Prior, which is used to test different dark energy models. Using this method, we compute the likelihood of each set of parameters $(\Omega_bh^2, \Omega_ch^2, h)$ and the corresponding cosmic age of the two models respectively, and figure out the best-fit value of the result. For the calculation of the uncertainties, we generated Monte Carlo Markov Chain.

Another advantage of selecting the age of the universe as a touchstone of distinguishing these two models should be mentioned here. The age of some of the old objects in the universe, such as metal-deficient stars, old galaxies, global clusters and etc., provide the lower limit of the age of the universe. By comparing the results, we can verify the validity of this specific model. Formed shortly after the Big Bang, metal-deficient stars are considered to share the same age as the universe roughly, and that is the reason why we choose HD 140283 \cite{HD} and HE 15230901 \cite{HE} for validity testing.

This paper is organized as follows. In section \ref{sec:model}, we review the LTB models and the reconstruction of this model constrained by the observations on the light cone. In section \ref{sec:calculation}, we introduce  the method to calculate the cosmic age in both in the $\Lambda$CDM model and the LTB-$\Lambda$CDM model, then we make use of the data given by 9-year WMAP \cite{WMAP9} to calculate the cosmic age in these two models and make likelihood analysis by generating Monte Carlo Markov Chain. In section \ref{sec:discussion}, we show the result of the parameters
 , verify the validity of these two models by comparing the results with the age of old objects in the universe and introduce the cosmic age method set in this paper. Finally, we give conclusions in section \ref{sec:conclusion}.

\section{Modeling}
\label{sec:model}
In subsection \ref{subsec:LTB}, we briefly review the Lema\^{\i}tre-Tolman-Bondi (LTB) model in general. Then in subsection  \ref{subsec:reconstruction}, following the procedures of \cite{CBKH}, we elucidate how to reconstruct a particular LTB model that exactly reproduces the selected observable features, namely the angular-diameter-distance-redshift relationship $d_A(z)$ and Hubble parameter $H(z)$, of the $\Lambda$CDM model.

\subsection{Lema\^{i}tre-Tolman-Bondi models}\label{subsec:LTB}
The LTB models are spherically symmetric cosmological solutions to the Einstein equations where the gravitational source is dust. Assuming that the system has purely radial motion and the motion is geodesic without shell crossing (otherwise we cannot ignore the pressure), the line element in the comoving and synchronous gauge can be written as:
\begin{equation}
\label{metric}
ds^2=-dt^2+\frac {R^{\prime 2}(r,t)} {1+2E(r)} dr^2 +
R^2(r,t)d\Omega^2,
\end{equation}
where $d\Omega^2=d\theta^2+\sin^2{\theta}d\phi^2$.
Here $R(r,t)$ represents the areal radius. The proper area of a sphere of coordinate radius $r$ on a time slice of constant $t$ is $4\pi R^2$. $E(r)$ plays two roles in LTB metric: (1) the geometric role, determining the local `embedding angle' of spatial slices that represents the spatial curvature; (2) the dynamic role, determining the local energy per unit mass of dust particles, hence the type of evolution of $R$, in other words, representing the energy per unit mass of the particles on that shell. The prime superscript is to denote $\partial/\partial r$, and the overdot to denote $\partial/\partial t$, in agreement with \cite{MHE}\cite{CBKH}. The Robertson-Walker metric can be recovered by performing $R(r,t)\rightarrow a(t)r$ and $2E(r)\rightarrow -kr^2$.

In spherically symmetric models, in general, there are two expansion rates:
\begin{eqnarray}
H_\perp&\equiv& \dot{R}(r,t)/R(r,t),
\nonumber\\
H_\parallel&\equiv& \dot{R}'(r,t)/R'(r,t),
\end{eqnarray}
at the transverse direction and the longitude direction respectively. In LTB models, the longitudinal expansion rate $H_\parallel(z)$ has the same form as Hubble parameter $H(z)$ in $\Lambda$CDM model, which is assured by definition\cite{LTB}.

With the dust equation of state, the Einstein field equations can be expressed as
\begin{equation}
\label{Einstein1}
H_\perp^2(r,t) + 2H_\parallel(r,t)H_\perp(r,t) -\frac{2E(r)}{R^2(r,t)}
- \frac{2E'(r)}{R(r,t)R'(r,t)} = \kappa\rho _M(r,t),
\end{equation}
\begin{equation}
\label{Einstein2}
\dot{R}^2(r,t)+2R(r,t)\ddot{R}(r,t)-2E(r)=0,
\end{equation}
where $\kappa=8\pi G$. These represent the generalization of the Friedmann equation for a homogeneous and isotropic universe to a spherically symmetric inhomogeneous universe.
Solving the Einstein equations, we get
\begin{equation}
\label{solution1}
H_\perp^2=\frac{\dot{R}^2(r,t)}{R^2(r,t)}=\frac{ 2E(r)}{R^2(r,t)} + \frac{2M(r)}{R^3(r,t)},
\end{equation}
\begin{equation}
\label{solution2}
\kappa\rho_M(r,t) = \frac{2M'(r)}{R^{2}(r,t)R^{\prime}(r,t)}.
\end{equation}
eq.(\ref{solution1}) can be solved in terms of a parameter $\eta=\eta(t,r)$:
\begin{eqnarray}
\label{solutiona}
R(r,t)&=&\frac{M(r)}{\chi(r)}\phi(r,t),\\
\label{solutionb}
t(r)-t_{\rm BB}(r)&=&\frac{M(r)}{\chi^{3/2}(r)}\xi(r,t),
\end{eqnarray}
where
\begin{equation}
\label{solutionc}
\chi(r)=\left\{
\begin{array} {ll}
2E(r) \\ 1 \\ -2E(r)
\end{array} \right.
~\phi=\left\{
\begin{array} {ll}
\cosh \eta-1
\\ \eta^2 /2
\\ 1-\cos \eta
\end{array} \right.
~\xi=\left\{
\begin{array} {ll}
\sinh \eta-\eta
\\ \eta^3/6
\\ \eta -\sin \eta
\end{array} \right.
~\textrm{when} \left\{
\begin{array} {ll}
E>0 ~(\textrm{hyperbolic evolution})
\\E=0 ~(\textrm{parabolic evolution})
\\ E<0 ~(\textrm{elliptic evolution}).
\end{array} \right.
\end{equation}
The LTB model is defined by three arbitrary functions of coordinate radius $r$: $M(r)$, $E(r)$, $t_{\rm BB}(r)$. The role of $E(r)$ is stated in the explanation of the metric. $M(r)$, appearing as a ``constant" while integrating eq.(\ref{Einstein2}), is the effective gravitational mass with comoving radius $r$, which characterizes the gravitational mass contained with the comoving spherical shell at any given $r$. $t_{\rm BB}$ is another arbitrary function that also comes out as an integration ``constant" and is interpreted as the ``Big Bang time". Note that there is no implication of a simultaneous bang surface without further assumptions.

We now denote quantities on the light cone by a hat. On radial null geodesics, $ds^2=d\theta^2=d\phi^2=0$. From eq.(\ref{metric}), the photon radial null geodesic equation for $\hat{t}(r)$ satisfies:
\begin{equation}
\label{tlc}
\frac{d\hat{t}(r)}{dr}=-\frac{R'(r,\hat{t}(r))}{\sqrt{1+2E(r)}}.
\end{equation}
where radial coordinate $r$ has no physical importance. In order to simplify calculation, we rescale $r$ on the light cone as
\begin{equation}
\label{rescale}
\widehat{R^\prime}=\sqrt{1+2E(r)}.
\end{equation}
The total derivative of $R$ is formed as:
\begin{equation}
\label{totdR}
\frac{d\widehat{R}}{dr}=\widehat{R'}+\widehat{\dot{R}}\frac{d\hat{t}}{dr}.
\end{equation}

At the origin of spherical coordinates ($r=0$), we assume that $R(0,t)=0$ and $~\dot{R}(0,t)=0$ for all $t$; the density is non-zero; the type of time evolution(hyperbolic, parabolic, elliptical) is consistent with its nearest neighborhood; all functions are smooth and have first derivatives.
Eq.(\ref{solutiona}) tell us that ${R(r,t)E(r)}/{M(r)}$ and ${E(r)^{3/2}}/{M(r)}$ must be finite at $r=0$. We expect $M\rightarrow 0$,when $r\rightarrow 0$, therefore $E\rightarrow0$ and $E\sim M^{2/3}$. Using the eqs.(\ref{solution1}) and(\ref{totdR}), we get:
\begin{equation}
\label{R0}
\frac{d\widehat{R}}{dr}\biggr\vert_{r=0}=\widehat{R'}\biggr\vert_{r=0}+
\widehat{\dot{R}}\frac{d\hat{t}}{dr}\biggr\vert_{r=0} = 1,
\qquad \widehat{R}=r,
\end{equation}
to the leading order.
Substitute this relationship into eq.(\ref{solution2}), we get
\begin{eqnarray}
\label{M0}
2M'(r)\vert_{r=0}&=&\frac{1}{2}\kappa \rho_M(r,t)R^2(r,t)R'(r,t)\biggr\vert_{r=0}\nonumber\\
&=& 3\Omega_M H_0^2 r^2,
\\2M(r)\vert_{r=0}&=&\Omega_m H_0^2 r^3,
\end{eqnarray}
to the leading order, where $\kappa \rho_{M0}=3\Omega_MH_0^2$ and $H_0$ is the Hubble constant.
And $E(r)$ satisfies:
\begin{eqnarray}
\label{E0}
H_{0}^2=\frac{2M(r)}{R^3}+\frac{2E(r)}{R^2},
\\2E(r)=(1-\Omega_m)r^2.
\end{eqnarray}

To match the theory with the observations, we need to associate the physical quantities with the redshifts, which is done by using the redshift equation \cite{zeq}
\begin{equation}
\frac{d \ln(1+z)}{dt}=-\frac{\dot{R}^{\prime}(r,t)}{R^{\prime}(r,t)}.
\end{equation}
Thus the redshift of the photon ${z}(r)$ takes the form
\begin{equation}
\label{zlc}
\frac{d{z}(r)}{dr}=(1+z)\frac{\dot{R}'(r, \hat{t}(r))}{\sqrt{1+2E(r)}}.
\end{equation}
Using the reciprocity theorem \cite{dadl}, the luminosity distance can be converted to the angular diameter distance:
\begin{equation}
\label{dlz}
\hat{d}_A(z)=\widehat{R}(z)=\frac{\hat{d}_L(z)}{(1+z)^2}.
\end{equation}

\subsection{Reconstructing LTB model with $\Lambda$CDM observational features on the light cone.}
\label{subsec:reconstruction}
This reconstruction procedure and results follow \cite{CBKH}, which allows one to construct an LTB model that reproduces: (1) the angular-diameter-distance-redshift relationship $\hat{d}_A(z)$; (2) the Hubble parameter $H(z)$ of the fiducial $\Lambda$CDM model. The reconstruction of LTB model was first set by \cite{MHE} to construct cosmological models that can fit the observations limited on the light cone with given source evolution, namely the absolute luminosity of the source at the time of emission $\widehat{L}(z)$ and true density over the source number density $\hat{m}(z)$. This theory was applied to reproduce the observables that match $\Lambda$CDM predictions by \cite{CBKH}. Note here that the profiles in this paper are result of reproducing observables that match $\Lambda$CDM model predictions, not from fitting to any real data.

The reconstruction procedures are listed as follows:
\begin{enumerate}
\item Use the first assumption that in the reconstructed LTB model, the angular-diameter-distance-redshift  relationship matches that of the $\Lambda$CDM model, where the angular diameter distance takes the form:
\begin{equation}
\widehat{d}_A(z) = \frac{1}{(1+z)} \int_0^z \frac{dz_1}{H_{\Lambda\textrm{CDM}}(z_1)} =\widehat{R}(z).
\end{equation}

\item Use the second assumption that in the LTB-$\Lambda$CDM model,
the Hubble parameter $H(z)$ is in agreement with the fiducial $\Lambda$CDM model, we get the relationship of $z$ and $r$.
\begin{eqnarray}
\label{zr}
\frac{dz}{dr}  & = &  (1+z) H_{\Lambda\textrm{CDM}}(z) {}\nonumber\\
& = & \left[ \frac{ d\widehat{R}(z)}{dz}(1+z)\right]^{-1}
\left[ 1- \frac{1}{2}\int_0^z \kappa\widehat{\rho}(z)\widehat{R}(z_1)
(1+z_1) \frac{dr}{dz_1}dz_1 \right ],
\end{eqnarray}
where the first equation is the derivation of Hubble parameter and the second equation is derived in the process of the reconstruction of the model(for more detailed formulae, see \cite{MHE} and \cite{CBKH}).

\item Solve the differential equation given by \cite{MHE} of $M(r)$ with an initial condition of $M(0)=0$,
\begin{equation}
\label{alpha}
\frac{dM}{dr}+\frac{\kappa \widehat{\rho}\widehat{R}}
{2 d\widehat{R}/ dr}M = \frac{\kappa \widehat{\rho}{\widehat{R}}^2}{4d\widehat{R}/dr}
\left[\left(\frac{d\widehat{R}}{dr}\right)^2+1 \right],
\end{equation}
And $E(r)$ is formed as:
\begin{equation}
\label{beta}
2E(r) = \left\{ \frac{1}{2}\left[\left(\frac{d\widehat{R}}{dr}\right)^2+1 \right]-\frac{M}{\widehat{R}}\right\}^2\biggr/ \left(\frac{d\widehat{R}}{dr}\right)^2-1
\end{equation}
\item Derive the third arbitrary function $t_{\rm{BB}}(r)$ of LTB-$\Lambda$CDM model that satisfies eqs.(\ref{solutiona}) and (\ref{solutionb}).
\end{enumerate}

\section{The calculation of cosmic age}
\label{sec:calculation}

\subsection{Cosmic age in $\Lambda$CDM model and LTB-$\Lambda$CDM model}
\label{subsec:age}

In the fiducial $\Lambda$CDM model, as mentioned in the introduction section, we treat the additional opponent as the time independent vacuum energy, a cosmological constant $\Lambda$. Relating the Hubble parameter with its present day value and ignoring radiation component (which is taken into consideration at high redshifts), we get:
\begin{equation}
\label{H}
H^2(z)=H_0^2 \left[\Omega_{m}(1+z)^3+\Omega_{\Lambda}\right].
\end{equation}
With the fact that ${da}/{a}=H(z)dt=-{dz}/{1+z}$, which leads to ${dz}/{dt}=-H(z)(1+z)$, we integrate eq.(\ref{H})to determine the age of the universe at a given redshift $z$:
\begin{equation}
\label{ageLCDMz}
t_{\Lambda\textrm{CDM}}(z)=\int_z^\infty \frac{dz}{(1+z)H(z)}
\end{equation}

As for the LTB-$\Lambda$CDM model, instead of solving the ordinary differential equations of $M(r)$ and $E(r)$ by eqs.(\ref{alpha}) and (\ref{beta}), which can only be solved numerically(see \cite{CBKH}). Here we take use of an easier but still precise way. The cosmic age is referred to the local universe. Therefore, we can just take use of the original conditions($r=0$) illustrated in Sec.\ref{subsec:LTB} eqs.(\ref{R0}),(\ref{M0}) and (\ref{E0})by taking $R(r,t)$, $M(r)$ and $E(r)$ to the leading order:
\begin{equation}
\label{initial}
\widehat{R}=r,
\qquad M(r)=\frac{1}{2}\Omega_m H_0^2r^3,
\qquad E(r)=\frac{1-\Omega_m}{2}r^2.
\end{equation}
Substitute these forms into eq.(\ref{solutiona}), we can get the solution of the parameter $\eta$ as:
\begin{equation}
\label{eta}
\cosh \eta=\frac{2-\Omega_m}{\Omega_m}.
\end{equation}
With eq.(\ref{solutionb}), we obtain the age of the universe in the LTB-$\Lambda$CDM model in the local universe($r=0$) as
\begin{equation}
\label{ageLTB0}
t_0-t_{\rm{BB}}(0)=\frac{\Omega_m}{2(1-\Omega_m)^{3/2}H_0}[\sinh\eta-\eta].
\end{equation}
which is also the analytical solution of the integration when $z=0$(referring to the local universe):
\begin{equation}
\label{anaageLTBz}
t_{\textrm{LTB}-\Lambda\textrm{CDM}}(0)=\int_0^\infty \frac{dz}{(1+z)H_0\sqrt{\Omega_{m}(1+z)^3+(1-\Omega_{m})(1+z)^2}}.
\end{equation}
Change the lower limit of integral $0$ into $z$, we can obtain a form to calculate the cosmic age at a given redshift $z$. We will later employ this form to calculate the cosmic age at $z=1.55$, shown in fig.\ref{fig2} .

\subsection{WMAP parameters and likelihood analysis}
\label{subsec:WMAP}
Now we choose suitable parameters to calculate the age of the universe respectively in $\Lambda$CDM model and LTB-$\Lambda$CDM model. Let us first trace back to\cite{MHE}.
The initial purpose to set an LTB model that can be found to fit given set of source evolution is to determine the degree of inhomogeneity from the observations and given source functions. Thus we need methods of validating source evolution models that do not depend on assumptions of homogeneity. Deep cosmological surveys, which supply measurements at high redshifts, may provide a good constraint.
The purpose of this paper is to find a method to distinguish between the inhomogeneous cosmic model and the $\Lambda$CDM model. We consider the fact that the inhomogeneity in LTB model plays a role of mimicking dark energy in fiducial $\Lambda$CDM model.

The WMAP Collaborations has determined values for several parameters by mapping the CMBR. Using the parameters $\Omega_m$\footnote{In $\Lambda$CDM model, matter contains the physical baryon components and the physical cold dark matter components, which satisfies:$\Omega_m=\Omega_b+\Omega_c$.} and $h$\footnote{$H_0=100 h {\rm km s}^{-1} {\rm Mpc}^{-1}$.}directly in eq.(\ref{ageLTB0}) overestimates the uncertainty, for these two parameters are degenerate. 5-year WMAP observations: Cosmological Interpretation\cite{WMAP5} gave a method to derive the best-fit parameters in different dark energy models. In this paper, we briefly review the Distance Prior method, then we will use this method to compute the likelihood of each set of parameters $(\Omega_bh^2,\Omega_ch^2,h)$ and find the best-fit parameters.

CMBR measures two distance ratios: (1) the angular diameter distance to the decoupling epoch divided by the sound horizon size at the decoupling epoch, $d_A(z_\ast)/r_s (z_\ast)$; (2) the angular diameter distance to the decoupling epoch divided by the Hubble horizon size at the decoupling epoch, $d_A(z_\ast)H(z_\ast)/c$.
This consideration to constrain various cosmology models. We shall qualify the first distance ratio, $d_A(z_\ast)/r_s (z_\ast)$, by the ``acoustic scale", $l_A$, which is defined by
\begin{equation}
\label{lA}
l_A\equiv (1+z_\ast) \frac{\pi d_A(z_\ast)}{r_s(z_\ast)}.
\end{equation}
The second ratio is often called the ``shift parameter", given by
\begin{equation}
\label{shiftp}
R(z_\ast)\equiv \frac{\sqrt{\Omega_m{H_0}^2}}{c} (1+z_\ast) d_A(z_\ast).
\end{equation}

We give the 9-year WMAP\cite{WMAP9} constraints on ($l_A$, $R$, $z_\ast$) that are recommended as the WMAP distance priors for constraining cosmology models, for a given set of parameters $(\Omega_bh^2,\Omega_ch^2,h)$. Here is the brief prescription for using the WMAP distance priors:
\begin{enumerate}
\item
Compute the redshift at the decouple epoch $z_\ast$ with the equation given below \cite{decz}
\begin{eqnarray}
\label{zast}
z_\ast= 1048[1+0.00124(\Omega_bh^2)^{-0.738}][1+g_1(\Omega_mh^2)^{g_2}]
\nonumber  \\
\textrm{where} \qquad g_1=\frac{0.0783(\Omega_bh^2)^{-0.238}}{1+39.5(\Omega_bh^2)^{0.763}},
\qquad g_2=\frac{0.560}{1+21.1(\Omega_bh^2)^{1.81}}.
\end{eqnarray}

\item Compute the comoving sound horizon size $r_s(z_\ast)$, with the specific form of Hubble parameter $H(z)$ in this model:
\begin{equation}
\label{rs}
r_s(z)=\frac{c}{\sqrt{3}}\int_{0}^{1/(1+z)}\frac{da}{a^2H(a)\sqrt{1+(3\Omega_b/4\Omega_\gamma)a}}.
\end{equation}.

\item Obtain the expression of angular diameter $d_A(z)$ in this model. Calculate $l_A$ and $R$ respectively with eqs.(\ref{lA}) and (\ref{shiftp}).
\item Form a vector in the order of $x_i =(l_A, R,z_\ast)$. The observed data vector is written as:
\begin{equation}
d_i=({l_A}^{\textrm{WMAP}},R^{\textrm{WMAP}},{z_\ast}^{\textrm{WMAP}})=(302.40,~ 1.7246, ~1090.88),
\end{equation}
where we choose the values with maximum likelihood.
Compute the likelihood $\mathcal{L}$, which is given by:
\begin{equation}
{\chi^2}_{\textrm{WMAP}} \equiv -2\mathcal{L} =(x_i-d_i)^{\rm    T}(C^{-1})_{ij}(x_j-d_j).
\end{equation}
\end{enumerate}

Here the inverse covariance matrix\cite{WMAP9} is formed as Table \uppercase \expandafter {\romannumeral 1}

\begin{table}[!h]
\label{table}
\tabcolsep 0pt
\vspace*{-10pt}
\begin{center}
\def\temptablewidth{0.5\textwidth}
{\rule{\temptablewidth}{1pt}}
\begin{tabular*}{\temptablewidth}{@{\extracolsep{\fill}}cccc}
  &   $l_A$ & $R$ & $z_\ast$\\ \hline
    $l_A$ & 3.182 & 18.253 &-1.429\\
$R$ &{}& 11887.879 & -193.808\\
$z_\ast$ & {} & {} & 4.556\\
       \end{tabular*}
       {\rule{\temptablewidth}{1pt}}
       \end{center}
\vspace*{-18pt}
       \caption{Inverse covariance matrix for the WMAP distance priors}
       \end{table}

For any given set of parameters $(\Omega_bh^2,\Omega_ch^2,h)$, we can calculate the age of the universe in $\Lambda$CDM and LTB-$\Lambda$CDM models respectively, and get the likelihood of the set of parameters. The set of parameters that gives the largest likelihood will be chosen to compute the best-fit value of cosmic age.

It should be stressed that since the LTB-$\Lambda$CDM model is reconstructed by reproducing the observable features on the light cone. The angular diameter distance $d_A(z)$ is exactly that in $\Lambda$CDM by reconstruction assumption. The longitude expansion rate $H_\parallel(z)$ equals the Hubble parameter $H(z)$ in $\Lambda$CDM model is guaranteed by the input requirements used to construct the model, which leads to the same form to calculate $r_s(z_\ast)$. Therefore, when estimating the acoustic scale $l_A$, the shift parameter $R$ and the redshift of the decouple epoch $z_\ast$, we can just obtain the derived values from the calculation of the $\Lambda$CDM model, there is no need to impose other special constraints on that.

We generate Monte Carlo Markov Chain by using the calculated likelihood to simulate the data. In this paper, we employ the Metropolis-Hastings algorithm specifically.
\begin{enumerate}
\item Choose a candidate set of parameters $s_\ast$ at random from a proposal distribution
\item Accept the candidate set of parameters with the probability $A(s,s_\ast)$; otherwise, reject it.
For Gaussian distributed observables (we assume these observables to be so), the acceptance function is:
\begin{equation}
A(s,s_\ast)=\min\{1, \exp[-\chi^2(s_\ast)+\chi^2(s)]\}.
\end{equation}
\end{enumerate}

\section{Discussion}
\label{sec:discussion}

\subsection{Bringing tighter constraints on the parameters of the parameters}
\label{subsec:breakdegeneracy}
\begin{figure}[htb]
\label{fig1}
\includegraphics[width=.85\textwidth]{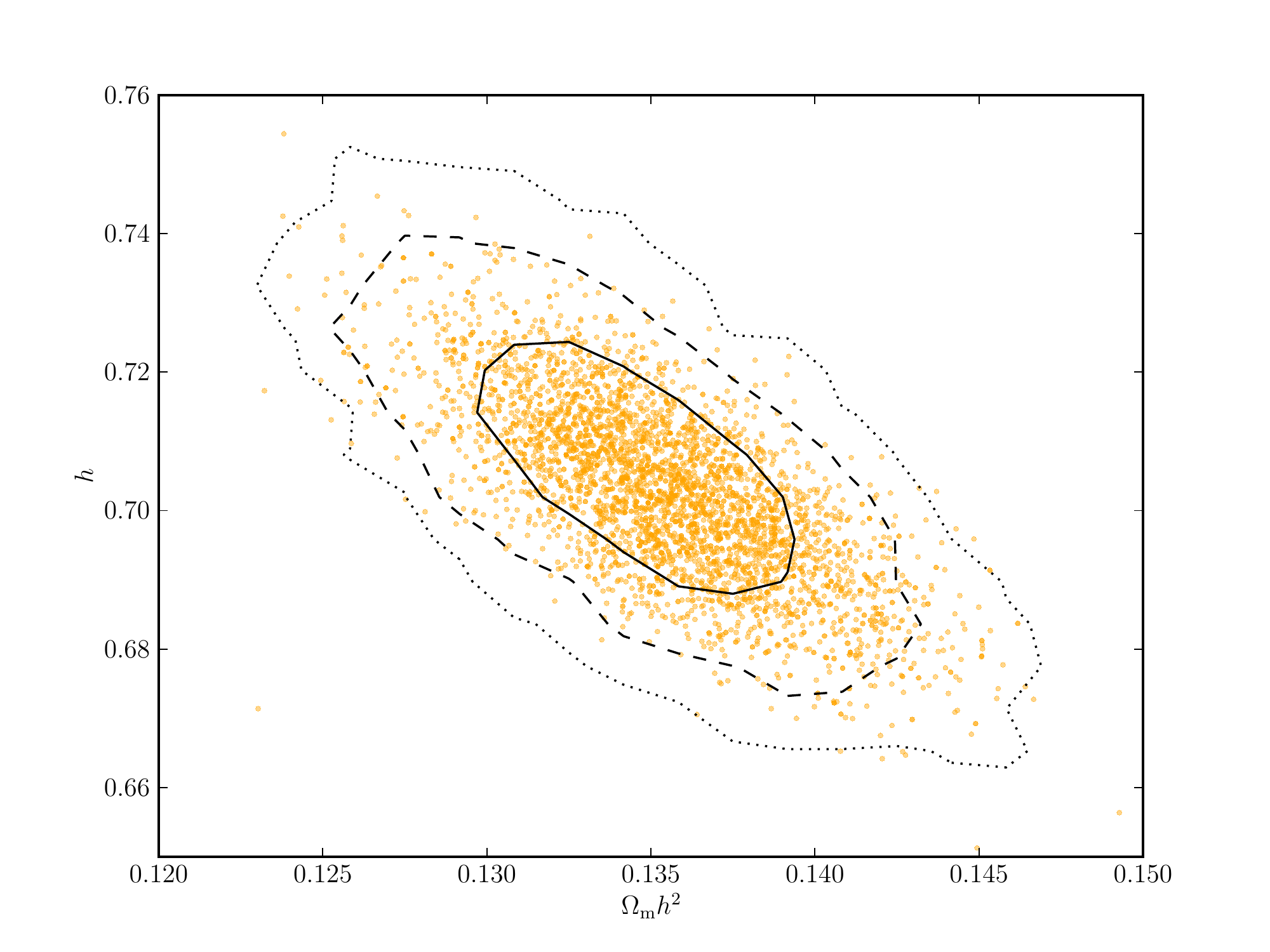}
\caption{Constraint on the parameters: $\Omega_m h^2$ and $h$. The contours show the $1\sigma$, $2\sigma$ and $3\sigma$ CL respectively from inside to outside.}
\end{figure}

\begin{figure}[htb]
	\includegraphics[width=.85\textwidth]{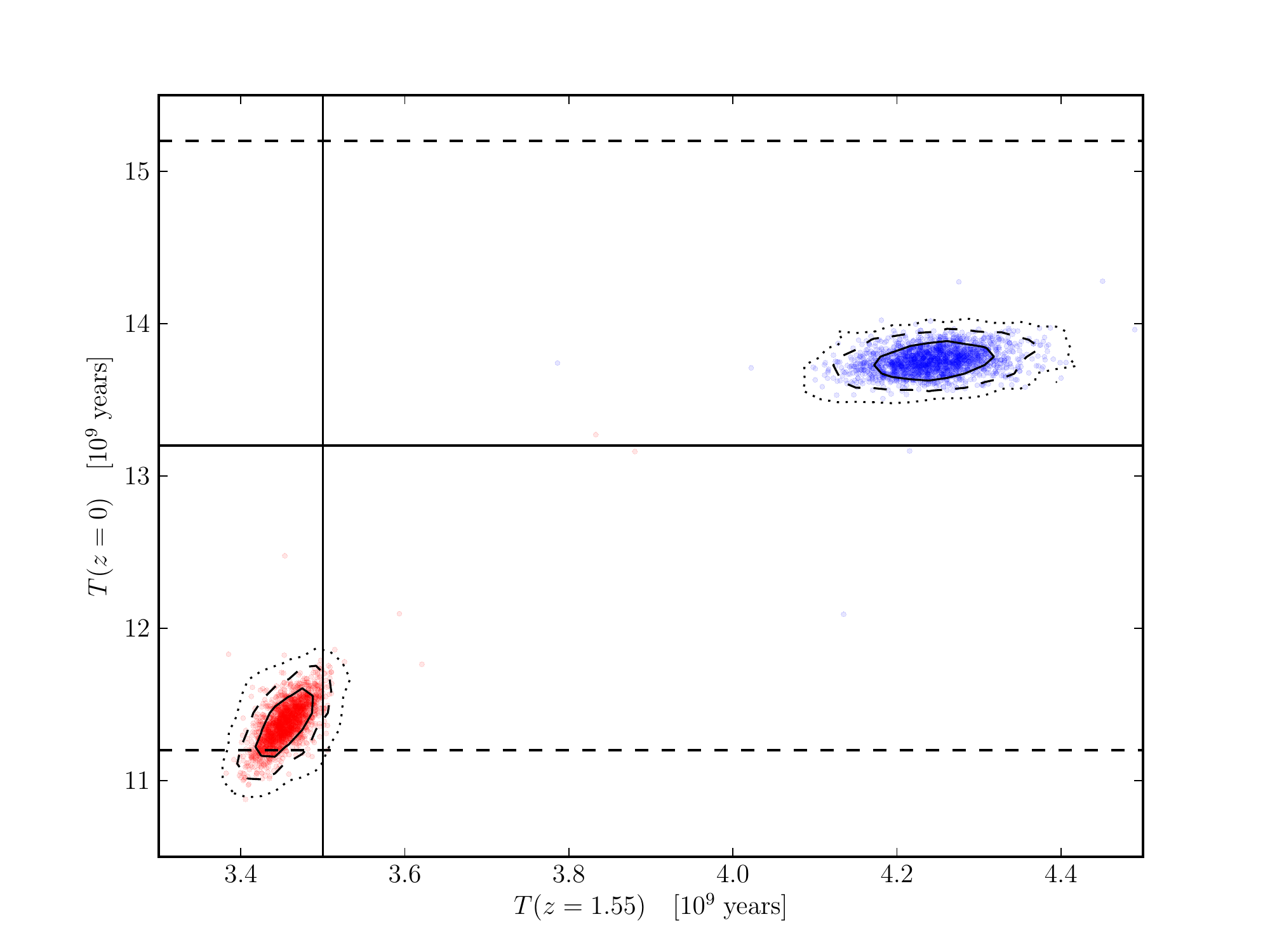}
	\caption{Constraints on the cosmic age at different redshifts. The horizontal and the vertical axis represent the cosmic age at $z=1.55$ and at $z=0$ respectively. The blue region plots the distribution of cosmic age for $\Lambda$CDM model, with the red region for  LTB-$\Lambda$CDM model. The contours show the $1\sigma$, $2\sigma$ and $3\sigma$ CL respectively from inside to outside. The horizontal solid line shows the age of metal-deficient star \cite{HE}, with the dashed lines showing the upper and lower limit of the stellar age in $1\sigma$ CL; the solid line in the vertical direction is the age of old galaxy at $z=1.55$ \cite{z=1.55}. Those lines provide a lower limit of the cosmic age at different redshifts.}
\label{fig2}
\end{figure}

WMAP 9\cite{WMAP9} provides the best-fit values of $\Omega_m$ and $H_0$ together with their $1\sigma$ confidence region. However directly using WMAP results will overestimate the uncertainty, for these two parameters are not orthogonal, in other words, their uncertainties are correlated. Computing the likelihood of any given set of parameters with Distance Prior method and generate Monte Carlo Markov Chain, we bring better constraints on the parameters. The result of Monte Carlo Markov Chain in parameter space is shown in figure\ref{fig1}.

The corresponding Monte Carlo Markov Chain of cosmic age results yields to be:
\begin{equation}
t_{\Lambda\textrm{CDM}} =13.76 \pm 0.09 ~\textrm{Gyr},
\qquad t_{\textrm{LTB}-\Lambda\textrm{CDM}} =11.38 \pm 0.15 ~\textrm{Gyr}.
\end{equation}

A similar previous work\cite{LLLW} should be mentioned here. They reconstruct LTB model by reproducing the luminosity distance $\hat{d}_L(z)$ and matter density $\widehat{\rho}(z)$ on the light cone, and compute the cosmic age in this model.\cite{LLLW} expand $R(r,t)$ near $r=0$, take this form to derive the approximation form of $M(r)$ and $E(r)$, and integrate eq.(\ref{solution1}) to obtain the expression of local cosmic age. Directly using the data given by 7-year WMAP \cite{WMAP7}, their results turn out to be :
\begin{equation}
t_{\Lambda\textrm{CDM}} =13.8 \pm 0.5 ~\textrm{Gyr},
\qquad t_{\textrm{LTB}-\Lambda\textrm{CDM}} =11.4 \pm 0.3 ~\textrm{Gyr},
\end{equation}
with the degeneracy of $\Omega_m$ and $H_0$ ignored. Although the models we use are based on reproducing different physical quantities limited on the light cone, the formulae that is used to calculate the age of the universe is quite the same, due to the same characters in the local universe. In comparison, the parameters we use yield more accurate results and less uncertainties.

Note that with the parameters given by PLANCK\cite{planck}, cosmic age in the standard model yield to be:
\begin{equation}
t_{\Lambda\textrm{CDM}} =13.81 \pm 0.058 ~\textrm{Gyr}.
\end{equation}

\subsection{Checking the validity of cosmic age}
\label{subsec:validity}
To assess the validity of the proposed models, we need to test the cosmic age in these models with a set of current observational data.
The lower limit of the cosmic age can be directly obtained from estimating the age of some old objects in our universe, such as the globular clusters, white dwarfs and metal-deficient stars. Formed shortly after the Big Bang, metal-deficient star is considered to be the more accurate lower limits of cosmic age among these ancient objects in the universe. The latest result of the age of a metal-deficient star in the solar neighborhood gives an age of $14.46 \pm 0.8$ Gyr\cite{HD}. Although the cosmic age in $\Lambda$CDM model is in consistence with its $1\sigma$ CL, it has quite large deviation from the age of the universe obtained from other calculations. Instead, we compare with a result that was published earlier, still of a metal-deficient star, giving the age of $13.2 \pm 2$ Gyr\cite{HE}, both the age of the two model are in $1\sigma$  consistence with that result.

Another cosmic age validity check is the age of an old galaxy at $z=1.55$\cite{z=1.55}, providing lower limit $3.5$Gyr of cosmic age at different redshifts. In fiducial $\Lambda$CDM model, the age at redshift $z$ can be calculated with eq.(\ref{ageLCDMz}). In LTB-$\Lambda$CDM model, with the origin conditions, we derive the way to calculate the cosmic age locally; however, eq.(\ref{anaageLTBz}) gives a formal method for age calculation at $z=1.55$, which can be applied for age validity check without caring about its physical meaning.

\subsection{Setting cosmic age method}
\label{subsec:theorem}

The method we use in this paper to calculate the age of the universe can be used as a set scheme for age calculation in different LTB models, which is based on the origin conditions of the spherical universe.
WMAP data can provide good constraints of age calculation. We need methods of validating source evolution models that do not depend on the assumptions of homogeneity to establish the age at any given $z$. Deep cosmological distance measures that is not influenced by source evolution would help to pin down the cosmological model better, the observations of CMBR in particular.
For cosmic age calculation in a certain LTB model, the prescription of this method is to be stated as follows:
\begin{enumerate}
\item Obtain the form of the light cone observables $\hat{d}_A(z)$ and $H(z)$ in this model.
\item Compute the age of this specific model using the method in section\ref{subsec:age}, and compute the likelihood following the prescription in section\ref{subsec:WMAP}. Figure out the best-fit value with the largest likelihood.
\item Generate the Monte Carlo Markov Chain with that likelihood(see section\ref{subsec:breakdegeneracy}), and make likelihood analysis using the chain.
\item Assess the validity of the age of the universe in this model by comparing the result with the age of the old objects in the universe. Here we recommend the age of metal-deficient stars.
\end{enumerate}

\section{Conclusion}
\label{sec:conclusion}
We derive the method to calculate cosmic age in $\textrm{LTB}-\Lambda\textrm{CDM}$ model with origin conditions of spherical coordinates. To bring tighter constraints on the parameters, we compute the likelihood of any given set of parameters in Distance Prior Method and generate Monte Carlo Markov Chain with that likelihood. The validity of the cosmic age is assessed by comparing the result with the age of the metal-deficient stars.

The observed normalization of the near-IR
galaxy luminosity function indicates that a void, if exists, amounts to a few hundred Mpc\cite{nearinfra}. Previous performing parameter estimation on the void model, the Hubble parameter data favor a void with characteristic radius of 2-3 Gpc. However, the test of such void models may ultimately lie in the future detection
of the discrepancy between longitudinal and transverse expansion rates, a touchstone of inhomogeneous models. Also, as pointed out in \cite{ksz}, one
can see that the Gpc-sized voids, as those favored by the supernovae data, are incompatible with the BIkSZ measurement, {are not favored}. However, as discussed in \cite{CBKH}, contrary to what is commonly claimed, LTB models with a giant void do not reproduce the main features of the $\Lambda$CDM model. These types of models just fit cosmological observations, with a priori constraints imposed on the LTB models.

It should be mentioned that in this paper we only consider a particular LTB model, specifically, the one reproducing the $\Lambda$CDM features limited on the light cone. Any statement about the universe off the light cone is unsupported by direct observations, and hence requires a cosmological model for the evolution of the universe. In this paper, besides $\Lambda$CDM model, we reconstruct an inhomogeneous LTB model by reproducing the angular-diameter-distance-redshift relationship $\hat{d}_A(z)$ and Hubble parameter $H(z)$. However, if one just restricted to the two observables used in this paper, it is impossible to uniquely determine a cosmological model. In our case, the fiducial $\Lambda$CDM model and the reconstructed LTB model result in identical observations for $\hat{d}_A(z)$ and $H(z)$. That result is valid even if one imagines perfect astronomical observations, since by construction the two cosmological observables are degenerate in the two models. In principle, whether it is possible to exclude the possibility just on the basis of light-cone observations is unclear now.
Cosmic age, which is off the light cone, can be a good choice to distinguish the reconstructed LTB model and the fiducial $\Lambda$CDM model. The method derived in this paper for age calculation and analysis can be used to a specific LTB model and by comparison with the observations to check the validity of this model.

This paper has developed and discussed issues related to the interpretation of cosmological observables in constructing cosmological models and theories. Observations of  $\hat{d}_A(z)$ and $H(z)$ do not prove that the so-called `dark energy' actually exists or that the universe is accelerating in the usual sense; they do so only if one assumes that the universe is homogeneous and isotropic as well as the dynamics of the universe are governed by General Relativity. The cosmic age method sheds light on a new way to distinguish the inhomogeneous model and  $\Lambda$CDM model.

\acknowledgments
Siqi Liu would like to acknowledge insightful suggestions given by Cong Ma, as well as useful discussions with Hao Wang, Shuo Yuan, and Hao-Ran Yu.
This work was supported by the National Science Foundation of
China (Grants No. 11173006), the Ministry of Science and Technology National Basic Science
program (project 973) under grant No. 2012CB821804, and the Fundamental Research Funds
for the Central Universities.



\begin{thebibliography}{99}
\bibitem{WMAP9}
Hinshaw G, Larson D, Komatsu E, et al., \emph{Nine-year Wilkinson Microwave Anisotropy Probe (WMAP) observations: cosmological parameter results}. arXiv preprint [arXiv:1212.5226, 2012].

\bibitem{SNR}
Riess A G, Filippenko A V, Challis P, et al. \emph{Observational evidence from supernovae for an accelerating universe and a cosmological constant} The Astronomical Journal, 1998, 116(3): 1009.

\bibitem{SNP}
Perlmutter S, Aldering G, Goldhaber G, et al. \emph{Measurements of $\Omega$ and $\Lambda$ from 42 high-redshift supernovae}. The Astrophysical Journal, 1999, 517(2): 565.

\bibitem{MHE}
Mustapha N, Hellaby C, Ellis G F R. \emph{Large-scale inhomogeneity versus source evolution: can we distinguish them observationally?}. Monthly Notices of the Royal Astronomical Society, 1997, 292(4): 817-830.

\bibitem{CBKH}
Celerier M N, Bolejko K, Krasinski A.  \emph{A (giant) void is not mandatory to explain away dark energy with a Lemaitre-Tolman model}. Astronomy and Astrophysics, 2010, 518.

\bibitem{WMAP5}
Komatsu E, Dunkley J, Nolta M R, et al.  \emph{Five-year wilkinson microwave anisotropy probe (WMAP) observations: cosmological interpretation}. arXiv preprint [arXiv:0803.0547, 2008].

\bibitem{HD}
	Bond H E, Nelan E P, VandenBerg D A, et al.  \emph{HD 140283: A Star in the Solar Neighborhood that Formed Shortly After the Big Bang} . The Astrophysical Journal Letters, 2013, 765(1): L12.

\bibitem{HE}
Frebel A, Christlieb N, Norris J E, et al.  \emph{Discovery of HE 1523–0901, a strongly r-process-enhanced metal-poor star with detected uranium}. The Astrophysical Journal Letters, 2007, 660(2): L117.

\bibitem{LTB}
Wang H, Zhang T J.  \emph{Constraints on Lemaître-Tolman-Bondi Models from Observational Hubble Parameter Data}. The Astrophysical Journal, 2012, 748(2): 111.

\bibitem{zeq}
Alnes H, Amarzguioui M,  \emph{Gr{\o}n {\O}. Inhomogeneous alternative to dark energy?}. Physical Review D, 2006, 73(8): 083519.

\bibitem{dadl}
Etherington I M H. LX.  \emph{On the definition of distance in general relativity}. The London, Edinburgh, and Dublin Philosophical Magazine and Journal of Science, 1933, 15(100): 761-773.

\bibitem{decz}
Hu W, Sugiyama N.  \emph{Small-scale cosmological perturbations: an analytic approach}. The Astrophysical Journal, 1996, 471(2): 542.

\bibitem{z=1.55}
Dunlop J, Peacockt J, Spinradi H, et al.  \emph{A 3.5-Gyr-old galaxy at redshift}. Nature, 1996, 381: 581.

\bibitem{LLLW}
Lan M X, Li M, Li X D, et al.  \emph{Cosmic age test in inhomogeneous cosmological models mimicking $\Lambda$CDM on the light cone.} Physical Review D[J], 2010, 82(2): 023516.


\bibitem{WMAP7}
Komatsu E, Smith K M, Dunkley J, et al.  \emph{Seven-year Wilkinson microwave anisotropy probe (WMAP) observations: cosmological interpretation}. The Astrophysical Journal Supplement Series, 2011, 192(2): 18.


\bibitem{ksz}
Zhang, Pengjie, and Albert Stebbins. "Confirmation of the Copernican Principle at Gpc Radial Scale and above from the Kinetic Sunyaev-Zel’dovich Effect Power Spectrum." Physical Review Letters 107.4 (2011): 041301.

\bibitem{nearinfra}
Keenan, R. C., A. J. Barger, L. L. Cowie, W-H. Wang, I. Wold, and L. Trouille. "Testing for a large local void by investigating the Near-Infrared Galaxy Luminosity Function." The Astrophysical Journal 754, no. 2 (2012): 131.

\bibitem{planck}
Ade, P. A. R., N. Aghanim, C. Armitage-Caplan, M. Arnaud, M. Ashdown, F. Atrio-Barandela, J. Aumont et al. "Planck 2013 results. XVI. Cosmological parameters." arXiv preprint arXiv:1303.5076 (2013).

\end{thebibliography}
\end{document}